\documentclass{article}
\usepackage{spconf,amsmath,graphicx,hyperref}
\usepackage{graphicx} 
\usepackage{subcaption}
\usepackage{float} 
\usepackage{amsmath}
\usepackage{bm}
\usepackage{booktabs} 
\usepackage{tabularx}

\title{WaveNeXt 2: ConvNeXt-Based Fast Neural Vocoders With \\Residual Denoising and Sub-Modeling for GAN and Diffusion Models}

\name{Wangzixi Zhou$^{\star \dagger}$, 
 Takuma Okamoto$^{\dagger}$, 
 Yamato Ohtani$^{\dagger}$, 
 Sakriani Sakti$^{\star}$,
 Hisashi Kawai$^{\dagger}$}
  
\address{$^{\star}$ Nara Institute of Science and Technology, Japan \\
      $^{\dagger}$ National Institute of Information and Communications Technology, Japan}

\begin{document}
\ninept
\maketitle

\begingroup
\renewcommand\thefootnote{}
\footnotetext{© 2026 IEEE. Personal use of this material is permitted. Permission from IEEE must be obtained for all other uses, in any current or future media, including reprinting/republishing this material for advertising or promotional purposes, creating new collective works, for resale or redistribution to servers or lists, or reuse of any copyrighted component of this work in other works.}
\endgroup

\begin{abstract}

Most neural vocoders are limited to one type: either GAN or diffusion-based. While state-of-the-art models like Vocos and WaveNeXt use powerful ConvNeXt-based generators, they have only been used in GAN frameworks and have limited performance in multi-speaker settings. Moreover, diffusion models, despite training faster than GANs, have slow CPU inference. In this paper, we introduce WaveNeXt 2, a unified ConvNeXt-based framework compatible with both GAN and diffusion vocoders. Its core innovation is residual denoising and sub-modeling, where each sub-model progressively refines the waveform. Experimental results in the multi-speaker dataset demonstrate the effectiveness of our approach: (1) GAN-WaveNeXt 2 is much faster than HiFi-GAN and WaveFit, and (2) Diff-WaveNeXt 2 also delivers much faster inference and competitive synthesis quality compared with FastDiff with 4 steps. The Diff-WaveNeXt 2 is very training-efficient, training in only 32 hours, making it ideal for resource-constrained applications.

\end{abstract}
\begin{keywords}
Vocoder, Diffusion, GAN, ConveNext, unified generator
\end{keywords}
\section{Introduction}
\label{sec:intro}

Neural vocoders have become fundamental components in modern speech synthesis systems, responsible for generating high-fidelity speech waveforms from acoustic features, such as mel-spectrograms. Compared with fast autoregressive (AR) models, non-AR models are stable and many models have been proposed. 
The fast and high-fidelity Neural vocoders are broadly categorized into two main architectures: generative adversarial network (GAN)- and denoising diffusion probabilistic model (DDPM)-based neural vocoders. Each approach offers distinct advantages and trade-offs in terms of synthesis quality, inference speed, and training complexity.

GAN-based models~\cite{melgan,yang2021multi,kong2020hifi,univnet,istftnet,bigvgan,ms-fc-hifi-gan,siuzdak2023vocos,okamoto2023wavenext}introduce a generator-discriminator framework to produce realistic waveforms with low latency. 
However, they often require substantial computational resources and are prone to instability during training. For example, training HiFi-GAN for 2.5 million steps can take over 300 hours on dual V100 GPUs. To mitigate these challenges, recent approaches, such as SpecDiff~\cite{baoueb2024specdiff}, integrate diffusion components to stabilize GAN training, while WaveFit~\cite{koizumi2023wavefit} replaces stochastic noise injection with a fixed-point strategy to guide the synthesis process more reliably.
In contrast, diffusion-based neural vocoders~\cite{kong2020diffwave,chen2020wavegrad,priorgrad,specgrad,huang2022fastdiff,fullband_diff,singing_diff,fastdiff2,
fregrad} leverage an iterative denoising process to generate speech waveforms by reversing a noise diffusion process. These models tend to be easier to train and more robust in certain scenarios but typically suffer from slow inference and potential degradation in output quality due to their multi-step generation pipeline. To improve inference efficiency, several techniques have been proposed to reduce the number of denoising steps. For instance, noise-level limited sub-modeling~\cite{okamoto2021noise} trains specialized sub-models for different noise ranges, enhancing prediction accuracy. BDDM~\cite{lam2022bddm} further accelerates inference by learning a compact noise schedule, enabling high-quality synthesis in as few as four steps.

While numerous methods have been proposed to accelerate inference speed, most fast neural vocoders are limited to either GAN or diffusion models, limiting flexibility in real-world applications. 
Inspired by ConvNeXt architectures originally developed for image processing~\cite{liu2022convnet}, recent generator designs in speech synthesis have attracted attention for their architectural simplicity and computational efficiency. 
Vocos \cite{siuzdak2023vocos} uses ConvNeXt blocks to predict STFT spectra, then reconstructs the waveform with an inverse STFT (iSTFT) layer. WaveNeXt \cite{okamoto2023wavenext} improves on this by using a trainable linear projection to directly predict the waveform, which enhances quality while maintaining speed.
However, these promising ConvNeXt-based generators have only been used in GAN-based frameworks. While they offer faster inference than models like HiFi-GAN, they still show limited performance in multi-speaker situations. This highlights the need for more versatile and robust solutions.

To realize fast and high-fidelity neural vocoders for GAN and diffusion models, we propose WaveNeXt 2, a unified ConvNeXt-based generator framework compatible with both diffusion and GAN vocoders. 
WaveNeXt 2 is the initial framework applicable to both GAN- and diffusion-based fast neural vocoders within a single architecture on a CPU.
\begin{itemize}
\item We introduce ConvNeXt-based residual denoising and sub-modeling, in which each sub-model gradually performs denoising at each time step in inference.
This enables a single architecture to be effectively applied across both vocoder types.
\item We achieve significant improvements in real-time factor (RTF): GAN-WaveNeXt 2 offers much faster inference with comparable quality to HiFi-GAN, WaveFit, and the original WaveNeXt, while Diff-WaveNeXt 2 achieves faster inference and competitive quality relative to FastDiff.
\end{itemize}
Speech samples from experiments are available on the demo page\footnote{https://37integer.github.io/WAVENEXT-2}.

\begin{figure*}[t]  
    \centering
    \begin{minipage}{0.48\textwidth}
        \centering
        \includegraphics[scale=0.4]{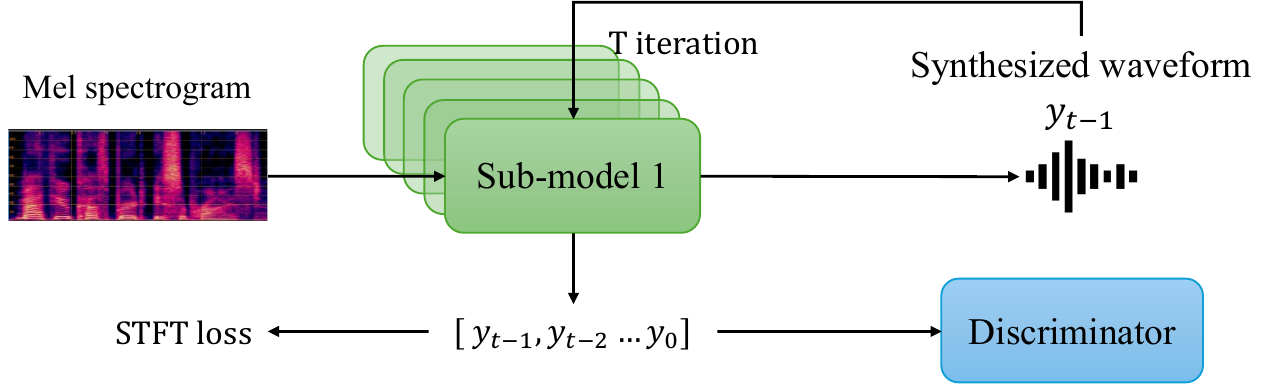}
        \vspace{0.035\textwidth}
        \subcaption{GAN-WaveNeXt 2}
        \label{fig:gan-based-2}
    \end{minipage}
    \hfill
    \begin{minipage}{0.48\textwidth}
        \centering
        \includegraphics[width=\textwidth, height=0.4\textwidth]{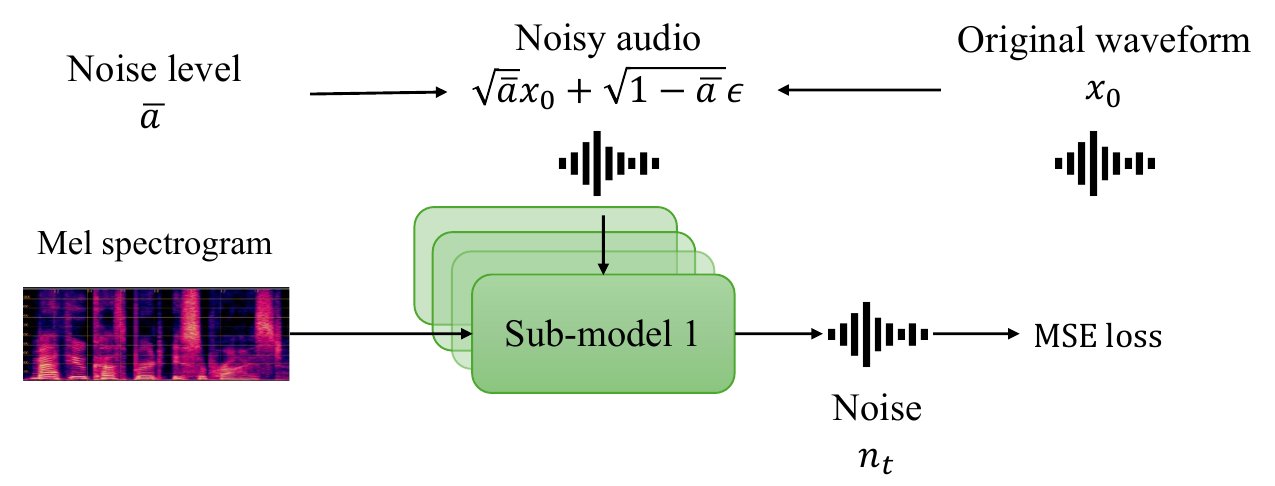}
        \subcaption{Diff-WaveNeXt 2}
        \label{fig:diffusion-based}
    \end{minipage}
    
    \caption{Training schemes of (a) GAN-WaveNeXt 2 and (b) Diff-WaveNeXt 2. In Diff-WaveNeXt 2, noise level $\overline{a}$ is predefined with the noise schedule predictor BDDM.}
    \label{fig:architecture}
    \vspace{-0.5cm}
\end{figure*}

\section{Related work} 
\label{sec:related work}

\subsection{Non-AR Neural Vocoders}

\noindent \textbf{GAN-based neural vocoders:} They use a generator and a discriminator in an adversarial process to produce high-quality speech. Models like HiFi-GAN \cite{kong2020hifi} and MS-FC-HiFi-GAN \cite{ms-fc-hifi-gan} have improved synthesis quality and inference speed through specialized architectures. Despite these advancements, GANs still require significant computational resources and can suffer from training instability. While newer methods like SpecDiff \cite{baoueb2024specdiff} and WaveFit \cite{koizumi2023wavefit} attempt to improve stability, GAN vocoders often have slow CPU inference speeds.

\noindent \textbf{Diffusion-based neural vocoders:} They generate speech by reversing an iterative denoising process. The need for many steps to achieve high quality slows down inference. To counter this, techniques have been proposed to reduce the number of steps. FastDiff \cite{huang2022fastdiff} uses specialized convolutions and a noise schedule predictor, while SpecGrad \cite{specgrad} employs an adaptive prior for better high-frequency quality. Additionally, noise-level limited sub-modeling \cite{okamoto2021noise} enhances prediction accuracy. However, even with these improvements, diffusion models still struggle with very few steps and their inference speed on CPUs remains slower than that of GANs.

\subsection{ConvNeXt in neural vocoders}

ConvNeXt~\cite{liu2022convnet}, originally introduced in the image domain, demonstrated remarkable accuracy while maintaining architectural simplicity and computational efficiency
. According to its outstanding performance, researchers have begun exploring its applications in the speech domain~\cite{siuzdak2023vocos,okamoto2023wavenext,convnext-tts}.

\noindent\textbf{Vocos}~\cite{siuzdak2023vocos} integrates ConvNeXt layers into a neural vocoder. It predicts high-resolution STFT spectra from input mel-spectrograms, which are then converted into waveforms using an iSTFT layer. Vocos achieves inference speeds up to ten times faster than HiFi-GAN on a CPU.

\noindent\textbf{WaveNeXt} \cite{okamoto2023wavenext} further improves upon this by replacing the iSTFT layer in Vocos with a trainable linear projection layer that directly predicts waveform samples, eliminating the need for spectral representations. This modification preserves the fast inference speed of Vocos while enhancing speech quality.

\section{Proposed approach}

We propose WaveNeXt 2, a unified framework that integrates ConvNeXt-based residual denoising sub-model into both GAN-based and diffusion-based architectures.

\subsection{ConvNeXt-based residual sub-modeling}

\begin{figure}[h] 
    \centering
        \begin{minipage}{0.22\textwidth}
        \centering
        \vspace{0.1\textwidth}
        \includegraphics[width=0.9\textwidth]{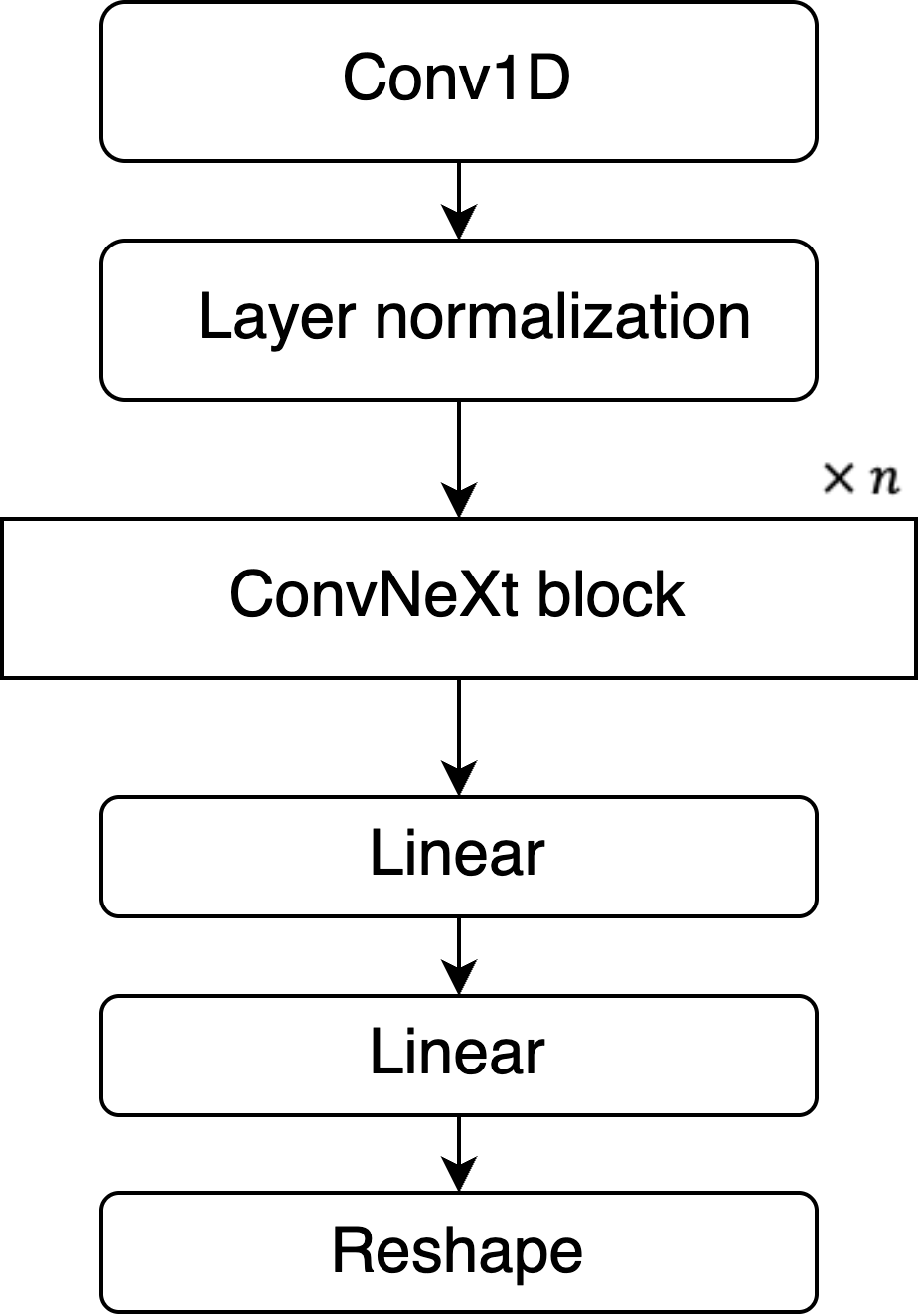}
        \vspace{0.05\textwidth}
        \subcaption{WaveNeXt-based generator}
        \label{fig:generator}
    \end{minipage}
    \hfill
    \begin{minipage}{0.22\textwidth}
        \centering
        \includegraphics[width=0.9\textwidth]{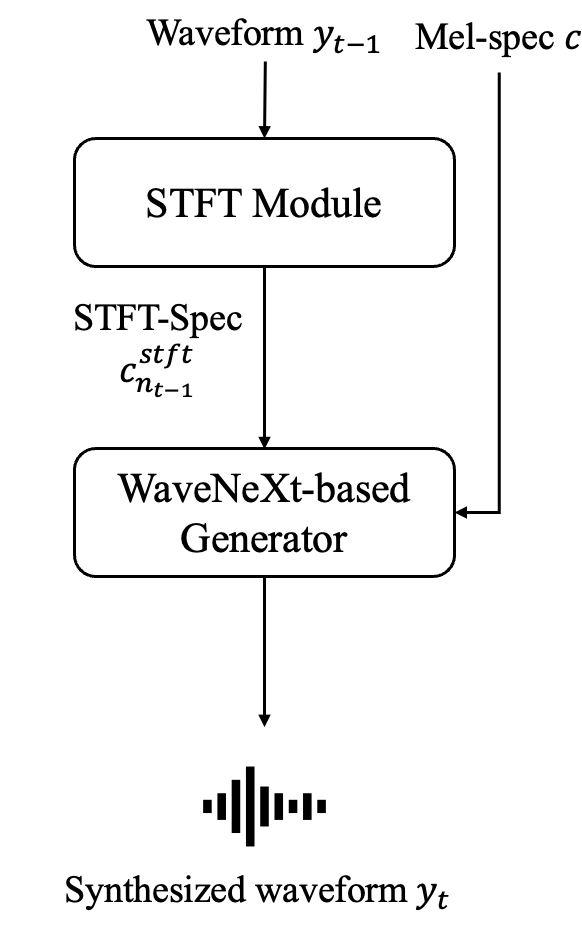}
        \subcaption{Sub-model}
        \label{fig:sub-model}
    \end{minipage}
    \caption{Overview of proposed architectures: (a) WaveNeXt-based generator where $n$ is the number of ConvNeXt blocks. $n=8$ is used in all the proposed models. (b) Sub-model for both GAN-WaveNeXt 2 and Diff-WaveNeXt 2, where $t$ is the number of iterations steps.}
    \label{fig:architecture}
    \vspace{-0.5cm}
\end{figure}

The architecture of the WaveNeXt-based generator is illustrated in Figure~\ref{fig:generator}. We retain the overall structure of the original WaveNeXt model~\cite{okamoto2023wavenext}, where the generator takes a mel-spectrogram as input and 

\hspace{-0.61cm}
directly outputs the synthesized speech signal \( y_0 \).
However, to enable a unified structure suitable for both GAN and diffusion frameworks, we modify the generator to predict the noise component \( n_t \) at each time step instead of generating the waveform directly. As shown in Fig.~\ref{fig:architecture}, the unified architecture comprises two main components: an STFT module and a WaveNeXt-based generator.

We first transform the input waveform \( y_{t-1} \) into its STFT representation using a Hann window. The STFT is computed with centering and produces a complex-valued spectrogram. The resulting complex spectrogram is then truncated along the temporal axis to match the duration of the target mel-spectrogram. The real and imaginary parts of the complex-valued STFT are separated for further processing.
To form a real-valued spectral representation compatible with the mel-spectrogram input, we concatenate the full real part of the STFT with the imaginary part excluding the DC and Nyquist components (i.e., omitting the first and last frequency bins). This results in an STFT-spec, which along with the mel-spectrogram, is fed into the WaveNeXt-based generator to predict the noise component \( n_{t-1} \) at the current time step.

\begin{figure}[h]
  \centering
  \centerline{\includegraphics[scale=0.42]{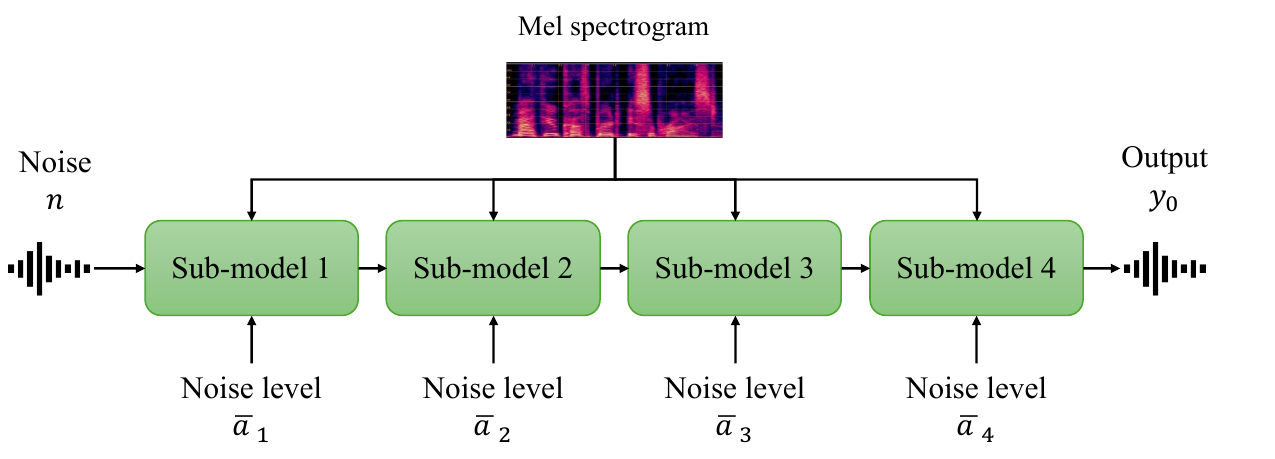}}
\caption{Inference procedure of Diff-WaveNeXt 2 with 4 sub models for 4 iterations.}
  \label{fig:inf}
\end{figure}

\subsection{GAN-based model: GAN-WaveNeXt 2}

The proposed GAN-WaveNeXt 2 is illustrated in Figure~\ref{fig:gan-based-2}. During training, we adopt the fixed-point iteration strategy introduced in WaveFit~\cite{9390397}, which differs from conventional DDPMs by deterministically guiding each denoising step toward the target waveform, rather than relying on stochastic noise removal.
The training process is structured as follows:  
In each iteration, the sub-model receives a mel-spectrogram and a noisy waveform \( y_t \) as input, and predicts the next denoised waveform \( y_{t-1} \). This process is repeated for \( T \) steps until the final waveform \( y_0 \) is synthesized. 

While our training pipeline is similar to WaveFit's, we have simplified it. WaveFit's original pipeline gradually converts initial input noise into clean speech based on fixed-point iteration, but we found that a "denoising" constraint is not necessary in the training loss, meaning initial input noise isn't required. We also omitted WaveFit's gain adjustment modules as they are redundant with the STFT loss. Preliminary experiments confirmed that GAN-WaveNeXt 2 performs effectively without either the initial input noise or the gain adjustment modules, allowing for a simplified training process.

\subsection{Diffusion-based model: Diff-WaveNeXt 2}
The architecture of the proposed Diff-WaveNeXt 2 is illustrated in Figure\ref{fig:diffusion-based}. 
Rather than following the original DDPM training strategy, we adopt the training strategy proposed in \cite{okamoto2021noise}, in which each sub-model is trained separately, each responsible for denoising within a specific range of noise levels. To implement this, we divide the denoising task into four stages and construct four sub-models. 
Each sub-model is conditioned not only on the mel-spectrogram but also on a specific noisy audio, denoted by \( x_t = \sqrt{\overline{a_t}} x_0 + \sqrt{1 - \overline{a_t}} \epsilon \),
where \( \epsilon \) represents Gaussian noise,  \( x_0 \) is original clean waveform and \(\overline{a_t}\) is the cumulative noise level predicted for step \( t \) .

The inference process is illustrated in Figure~\ref{fig:inf}. The mel-spectrogram and the corresponding noise level \( \overline{a} \) are provided as inputs to the respective sub-models. Starting from an initial noise signal \( \boldsymbol{n} \), the four sub-models are applied sequentially, each responsible for denoising within a specific noise level range. After four sub-model, the final output is the synthesized speech waveform \( \boldsymbol{y}_0 \).
According to \cite{okamoto2021noise}, the high-frequency details in synthesized speech are often lost when the noise schedule contains unnecessary noise, especially with a low number of iterations. To restore these missing components, we also use the time-invariant spectral enhancement post-filtering technique introduced in the same paper.

\begin{table*}[t]
    \centering
    \footnotesize
    \caption{Results of mel-cepstral distortion (MCD), log F0 root-mean-square error (RMSE), UTMOS and NISQA columns represent the means and standard deviations. Real-time factor (RTF) of the inference. Proposed methods are described in {\bf bold}.}
    \label{tab:table1}
     \begin{tabular}{lccccccc}
        \toprule
        & RTF(GPU) $\downarrow$ & RTF(CPU) $\downarrow$ & NISQA $\uparrow$ & UTMOS $\uparrow$ & MCD $\downarrow$ & log F0 RMSE $\downarrow$ & Model size  \\
        \midrule
        Ground Truth & -- & -- & 4.08 ± 0.19& 4.11 ± 0.09& -- & -- \\
        \midrule
        WaveNeXt (1 iteration) & \textbf{0.0022}& \textbf{0.06}& 3.16 ± 0.24 & 3.20 ± 0.12& 0.92 ± 0.52& 0.31 ± 0.15&14.98M\\
        WaveFit (2 iterations) & 0.0111 & 2.15 & 3.80 ± 0.22& 3.89 ± 0.11& 1.03 ± 0.54& 0.32 ± 0.15&15.51M\\
        \textbf{GAN-WaveNeXt 2 (2 iterations)} & 0.0033 & 0.10& 3.77 ± 0.20& 3.88 ± 0.11& 0.97 ± 0.54& 0.31 ± 0.15&29.97M\\
        WaveFit  (3 iterations)& 0.0151 & 3.22 & 3.91 ± 0.22& 3.98 ± 0.10& 1.01 ± 0.54& 0.32 ± 0.13&15.51M\\
        \textbf{GAN-WaveNeXt 2 (3 iterations)} & 0.0054 & 0.15 & 3.92 ± 0.22& 3.91 ± 0.10& 0.96 ± 0.57& 0.30 ± 0.18&44.96M\\
        WaveFit (4 iterations) & 0.0213 & 4.28 & 3.97 ± 0.21& 3.99 ± 0.10& 1.01 ± 0.52& 0.32 ± 0.11&15.51M\\
        \textbf{GAN-WaveNeXt 2 (4 iterations)} & 0.0066 & 0.20 & 4.01 ± 0.20& 4.04 ± 0.09& 0.95 ± 0.53& 0.30 ± 0.11&59.94M\\
        WaveFit (5 iterations) & 0.0226 & 5.36 & 4.02 ± 0.19 & 4.04 ± 0.09& \textbf{0.90 ± 0.52}& 0.31 ± 0.13&15.51M\\
        \textbf{GAN-WaveNeXt 2 (5 iterations)} & 0.0090 & 0.24 & 4.01 ± 0.19 & 4.04 ± 0.09& 0.95 ± 0.51& 0.30 ± 0.12&74.93M\\
        HiFi-GAN V1& 0.0110& 0.80 & 3.99 ± 0.22& \textbf{4.05 ± 0.11}& 2.34 ± 0.83& 0.16 ± 0. 01&13.9M \\
        \midrule
        FastDiff wo/ sub-model & 0.0625 & 0.80 &3.43 ± 0.20& 3.50 ± 0.11& 4.76  ± 0. 74& 0.16 ± 0. 01&15.63M\\
        \textbf{Diff-WaveNeXt 2 wo/ sub-model} & 0.0335& 0.16&3.45 ± 0.19& 3.55 ± 0.09& 7.34  ± 1. 46& 0.16 ± 0. 01&14.42M\\
        FastDiff w/ sub-model& 0.0282 & 0.80 & 3.67 ± 0.20& 3.78 ± 0.06& 4.32 ± 0.69& 0.24 ± 0.33&62.52M\\
        \textbf{Diff-WaveNeXt 2} & 0.0164 & 0.16 & 3.81 ± 0.19& 3.87 ± 0.05& 4.16  ± 0. 88& \textbf{0. 12 ± 0. 01}&57.68M\\
        \bottomrule
    \end{tabular}
\end{table*}

\section{Experiments} \label{seq:exp}
All the models were implemented using PyTorch ~\cite{pytorch} and trained on NVIDIA A100 GPUs with 40 GB of memory.


\begin{figure}[h]
  \centering
  \centerline{\includegraphics[scale=0.3]{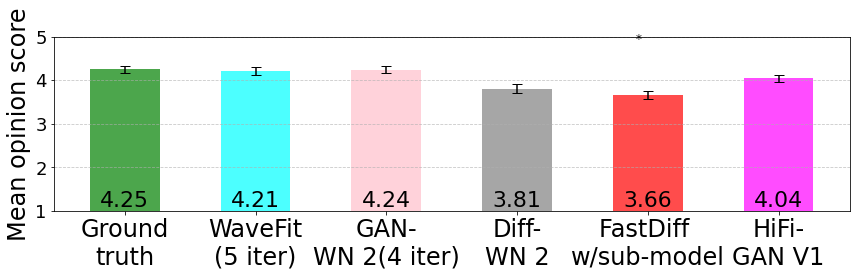}}
  \vspace{-0.2cm}
  \caption{Results of MOS tests with 20 listening subjects. The confidence level is 95\%. WN means WaveNeXt.}
  \label{fig:results}
\end{figure}

\subsection{Experimental conditions}
\textbf{Dataset}: We trained all the models on LibriTTS-R \cite{koizumi2023libritts}, which is a multi-speaker English corpus of approximately 585 hours of read English speech at 24~kHz sampling rate. We trained model from the combination of the “train-clean-100” and “train-clean-360” subsets at 24~kHz sampling. 

\noindent\textbf{Model and training setting}: 
For GAN-based models, we introduced unofficial implementations of HiFi-GAN V1\footnote{https://github.com/kan-bayashi/ParallelWaveGAN} and WaveFit\footnote{https://github.com/yukara-ikemiya/wavefit-pytorch}. For diffusion-based models, we introduced an official implementation of FastDiff\footnote{https://github.com/Rongjiehuang/FastDiff}. To implement WaveNeXt-based generator, we introduced an official implementation of Vocos\footnote{https://github.com/gemelo-ai/vocos} and replaced the STFT layer with linear layer.
For all the models, 128-dimensional mel-spectrograms were used as input acoustic features in WaveFit. To ensure a fair comparison with baseline systems, we matched the hop sizes of the proposed models to those of the corresponding baselines: GAN-WaveNeXt 2 and HiFi-GAN adopted a hop size of 300, consistent with WaveFit, while Diff-WaveNeXt 2 used a hop size of 256, matching that of FastDiff.
In GAN-WaveNeXt 2, we employed the same discriminators and loss functions as WaveFit to ensure training stability and consistency across adversarial learning. All losses follow the same definitions as those in WaveFit \cite{koizumi2023wavefit}. 
In Diff-WaveNeXt 2, we trained four independent sub-models to handle different noise levels. The noise schedule used for training was predicted by a noise schedule predictor adapted from BDDM~\cite{lam2022bddm}. The resulting noise schedule with 4 steps was [1.0e$-04$, 2.8e$-02$, 5.6e$-01$, 9.1e$-01$].

\noindent\textbf{Evaluation criteria}: To evaluate the naturalness of the synthesized speech subjectively, we conducted mean opinion score (MOS) tests~\cite{p800} using a five-point scale. 
A total of 20 paid native English speakers participated in the evaluation. All subjects use headphones in a quiet environment to listen. In total, each participant evaluated 120 samples (20 utterances $\times$ 6 models). 
In addition subjective evaluation, we employed objective evaluation methods to assess speech quality. We measured UTMOS~\cite{saeki2022utmos} and NISQA~\cite{mittag2021nisqa}, which are automatic MOS prediction models.
And we adopted two widely used signal-based objective metrics: mel-cepstral distortion (MCD)~\cite{407206} and log F0 root-mean-square error (RMSE), both of which provide quantitative assessments of spectral and prosodic accuracy. 
For inference speed, we measured RTFs on an NVIDIA A100 GPU and an AMD EPYC 7542 CPU (1 core).

\noindent\textbf{Ablation study}: For Diff-WaveNeXt 2, we also implemented our model within the original FastDiff architecture without sub-modeling. However, the performance was suboptimal compared to our unified framework. Therefore, we introduced sub-modeling train strategy~\cite{okamoto2021noise} to realize better performance. The complete ablation results and comparisons are summarized in Table~\ref{tab:table1}.

\begin{table}[h]
    \centering
    \caption{Training time of models in a single GPU}
    \vspace{-0.2cm}
    \label{tab:time}
    \begin{tabular}{l@{\hspace{2cm}}c} 
        \toprule
        Model & Training time \\
        \midrule
        \textbf{GAN-WaveNeXt 2} & 410 hours \\
        HiFi-GAN & 270 hours \\
        WaveFit & 410 hours \\
        \textbf{Diff-WaveNeXt 2} & \textbf{32 hours} \\
        Fastdiff & 96 hours \\
        \bottomrule
    \end{tabular}
    \vspace{-0.5cm}
\end{table}

\subsection{Results and discussion}
\label{sec:result}

We evaluated the performance of our proposed models in terms of both the inference speed and synthesized speech quality. The RTFs on a GPU and a CPU, as well as objective speech quality metrics—log F0 RMSE, Mel Cepstral Distortion (MCD), UTMOS, and NISQA—are shown in Table~\ref{tab:table1}. The evaluations were conducted using  4,824 samples from the LibriTTS-R \cite{koizumi2023libritts} “test-clean-100” subset. Additionally, the results of MOS tests obtained from 20 samples of the same subset are used to assess subjective quality. These are shown in Figure~\ref{fig:results}.

The results demonstrate that GAN-WaveNeXt 2 achieves UTMOS and NISQA scores comparable to those of WaveFit, while drastically improving the inference speed. Specifically, GAN-WaveNeXt 2 achieves a 70\% reduction in RTF on GPU and a 90\% reduction on CPU, compared to WaveFit. As illustrated in Figure~\ref{fig:results}, GAN-WaveNeXt 2 with 4 iterations achieves comparable MOS scores with both WaveFit with 5 iterations and HiFi-GAN, while also surpassing HiFi-GAN in terms of the inference speed—offering a 40\% improvement on GPU and 75\% on CPU. Although GAN-WaveNeXt 2 performs higher log F0 RMSE compared to HiFi-GAN, it performs better in terms of MCD, indicating superior spectral fidelity.

For the diffusion-based models, Diff-WaveNeXt 2 also produces synthesized speech quality comparable to that of FastDiff. By adopting the fixed noise conditioned sub-model training strategy proposed in~\cite{okamoto2021noise}, the speech quality is further improved. Notably, Diff-WaveNeXt 2 with sub-modeling achieves lower log F0 RMSE than HiFi-GAN.
Compared to FastDiff, Diff-WaveNeXt 2 offers a 36\% reduction in RTF on GPU and an 80\% reduction on CPU, significantly enhancing inference efficiency.

All of the models' training times are detailed in Table~\ref{tab:time}.
GAN-based models generally require substantial computational resources to train to convergence. For example, HiFi-GAN takes around 270 hours, while WaveFit and our proposed GAN-WaveNeXt 2 both require approximately 410 hours.
In contrast, diffusion-based models are significantly more efficient. Our proposed Diff-WaveNeXt 2 requires only 32 hours of training, a major reduction compared to the 96 hours typically needed for FastDiff. This notable decrease in computational and time costs makes Diff-WaveNeXt 2 highly suitable for large-scale or resource-constrained applications, as it maintains acceptable speech synthesis quality with a much lighter training burden.

Although sub-modeling improves training efficiency and performance, it also increases the total size of the model, as described in Table \ref{tab:table1}. The overall parameters will grow with the number of sub-models. This is an issue of the proposed methods.

\section{Conclusion}
This paper introduces WaveNeXt 2, a unified ConvNeXt-based generator with a residual denoising sub-modeling structure that is the first to be compatible with both GAN- and diffusion-based neural vocoders. Our approach successfully addresses the performance limitations of the original GAN-based WaveNeXt in multi-speaker scenarios, where it previously underperformed compared to HiFi-GAN. By extending the use of ConvNeXt-based generators beyond the GAN framework, WaveNeXt 2 allows for a direct, intuitive comparison between the two major architectures. Our results demonstrate that both GAN-WaveNeXt 2 and Diff-WaveNeXt 2 achieve high performance. Specifically, GAN-WaveNeXt 2 provides significantly faster inference, especially on a CPU, while maintaining synthesis quality comparable to HiFi-GAN and WaveFit. Concurrently, Diff-WaveNeXt 2 delivers much faster CPU inference and superior perceptual quality compared to a 4-step FastDiff model, effectively outperforming it in both efficiency and quality. This unified framework offers flexible choices for various applications: for fast deployment in resource-constrained environments, Diff-WaveNeXt 2 is the ideal choice, whereas for scenarios demanding the highest synthesis quality, GAN-WaveNeXt 2 is the better option.

\vspace{0.5cm}
\noindent\textbf{Acknowledgement:} 
Part of this work was supported by JSPS KAKENHI Grant Numbers JP21H05054, JP23K21681, JP24K0296, and JP25H01139, as well as JST NEXUS (JPMJNX25C1).
\vfill\pagebreak

\bibliographystyle{IEEEbib}
\bibliography{strings,refs}

\end{document}